\documentclass{aa}
\usepackage{natbib}

\usepackage{graphicx}
\usepackage{txfonts}

\begin{document}

\title{Cosmic voids and the kinetic analysis}
       \author{V.G.Gurzadyan\inst{1,2}, N.N.Fimin\inst{3}, V.M.Chechetkin\inst{3} }

              \institute{Center for Cosmology and Astrophysics, Alikhanian National
Laboratory and Yerevan State University, Yerevan, Armenia \and
SIA, Sapienza Universita di Roma, Rome, Italy \and Keldysh Institute of Applied Mathematics of RAS, Miusskaya Sq., 4, Moscow, Russia}

   \offprints{V.G. Gurzadyan, \email{gurzadyan@yerphi.am}}
   \date{Submitted: XXX; Accepted: XXX}

 \abstract{
The kinetic approach to the formation of the filaments in the large-scale matter distribution in the Universe is
considered within the Vlasov formalism. The structures arise due to the self-consistent dynamics, along with the repulsive term in
the modified Newtonian gravity which includes the cosmological constant.
That modified gravity enables one to describe the Hubble tension as a result of two flows, the local and global ones.
The criteria for formation of non--stationary semi-periodic structures in a system of gravitating particles described by Vlasov--Poisson equations are obtained in the case of that repulsive term.
The obtained dispersion relations for the Vlasov equation in the vicinity of the
singular point of the modified gravitational potential demonstrate
the possibility of the emergence of filaments as coherent complex states of relative
equilibrium  in non--stationary systems as structures of low dimensions (walls), and voids between them, of scales (diameters) defined by the balance between the gravity and repulsive term of the cosmological constant.}


   \authorrunning{N.N.Fimin, V.M.Chechetkin, V.G. Gurzadyan}
   \titlerunning{Filaments and the kinetic analysis}
   \maketitle
%

\section{Introduction }

The evolution of primordial density fluctuations is considered to be responsible for the observed large-scale matter 
distribution in the Universe (\cite{Peeb,BK}).
The assumptions on the spectrum of the initial fluctuations and analysis of various phases of the evolution of the 
perturbations enable one to cover the notable features of the observational surveys.

Zeldovich's pancake theory (\cite{Z,Arn,ArnP,SZ,SS}) is currently the basis of studies of the filament formation, 
including, including a direct comparison with observations.
Recently, the Hubble tension \citep{R,Val,R1}, that is the discrepancy between the local and global values of the Hubble parameter, and 
other observational tensions, appear to indicate the need to reconsider the views regarding the dynamics and 
structure at least in the local Universe.  The modified gravity approaches are actively studied among other options; see \cite{photon,Val}, and references therein. One 
of those approaches is based on a key principle of gravitational interaction being reconsidered, that is to say 
the equivalency of the gravity of the sphere and of the point mass located in its center. Namely, the theorem proved 
in \citet{G1} states that the general function satisfying that principle has the following form:
\begin{equation}
F(r)= Ar^{-2} + \Lambda r,
\end{equation}
where the first term in the r.h.s is Newton's law, while the second term defines the cosmological 
constant $\Lambda$ in the McCrea-Milne cosmological scheme (\cite{MM}) and it corresponds to the cosmological term 
in the solutions of Einstein equations and weak-field general relativity (GR) (\cite{GS2,GS3}).

In \citet{Z81}, the validity of the non-GR description of the local Universe was outlined. Eq.(1) enables one to 
describe the dynamics of groups and clusters of galaxies (\cite{GS2,G2,GS3,GS4}), that is at distance scales 
when the second term becomes non-negligible with respect to the Newtonian term.

Eq.(1) possesses a remarkable feature, namely, it defines a non-force-free field inside a spherical shell, 
thus differing from Newton's law when the shell does not influence its interior. In fact, there are 
observational indications that the galactic halos do indeed determine certain features of the spirals and disks (\cite{Kr}), 
thus supporting the non-force-free shell's concept. Eq.(1) enables one to describe the dynamics of groups 
and clusters of galaxies, and especially the Hubble tension, as a possible indication of two flows, a local one described 
by the McCrea-Milne nonrelativistic equation with a cosmological term and the global one, described by Friedmannian equations \cite{GS7,GS8}.

We study the role of law Eq.(1) within the Vlasov kinetic formalism \cite{V1,V2,VFC}, thus outlining the notable differences from previous approaches to filament formation. On the one hand, it is a continuation of Zeldovich's approach of the validity of the nonrelativistic 
treatment of the local structure formation. On the other hand, it also considers the role of the self-consistent 
gravitational interaction in that process, however now with a repulsive term of Eq.(1), thus continuing the studies in \cite{web}.

The modification of the Newtonian gravitational potential in Eq.(1) \cite{G1}
leads to an effective negative pressure in the system of interacting
particles. So the equilibrium condition for an element of matter, taking the repulsive effect of the $\Lambda$ term
into account, determines the scales of the formation of macro-structures in the Universe. 

Namely, at scales smaller than corresponding the one of the condition of balance of gravitational forces and the
repulsive $\Lambda$ term at the Jeans wavelength determined by the inflection point of potential in Eq.(1) (Fig.1),
the collapse of homogeneous matter occurs, and at larger scales fragmentation of the system occurs. 
From the scale of the inflection in the potential Eq.(1), we obtain the mean dimensions of the voids, that is around 25 Mpc, if oriented on the mean density in the Virgo Supercluster.

Another representation of the mutual balance of the gravity and repulsive term in Eq.(1) can be the abovementioned two-flow nature of the
Hubble flow \cite{GS7,GS8}. 

The next principal issue is the coherence of the relaxation of perturbations
in the system along the selected direction, that is to say at large structures --- along
two mutually perpendicular directions. Zeldovich's mechanism of coherence in Lagrangian coordinates \cite{Z,Arn}
takes the existence of synchronism of anisotropic damping of emerging density fluctuations into account, which is
natural within the framework of the theory of long-range interactions.
For a system of massive particles, their uniform distribution cannot be considered as a global equilibrium in
configuration space (this is implied by the divergence of the potential in the integral form of the Poisson equation). That is why
it is necessary to refer to the representation of the aforementioned
energy substitution through the ``minimum'' potential function of the gravitational
interactions $\Phi_{min}({\bf x})$ for a many--particle system with an
additional condition that singles out its singular point.

In a system of particles obeying the generalized law of Eq.(1), there are several types of
singular points in the configuration space (in the general case of its subdomains),
the main ones are the following:

1)\:libration points in a many-particle system;

2)\:inflection points of the generalized gravitational
potential of Eq(1); see Fig.(1).

Regarding the first type, we can definitively state that in the absence of a high degree of symmetry in the system, finding such points is practically not feasible (exceptions are only Lagrange points on Keplerian orbits), and their stability
requires the development of special methods, see  \cite{Sch}, for example.
Particularly, in \cite{web}, the solutions of the nonlinear
Liouville--Gelfand equation (LGE) \cite{Gelfand} are studied to describe the inter-particle potential in the stationary
version of the boundary value problem. This shows the need for special tools and a technique to reveal
the features of the structures that emerged as a result of the Vlasov treatment of the problem.

Below, we concentrate on analyzing the dispersion relations for Vlasov-Poisson equations,  corresponding to
conditionally equilibrium solutions, defining pseudo--ordered structures of low dimensions (walls),
semi-periodic structures and inter-structure regions (voids).

\section{The initial perturbation problem for the linearized integral Vlasov--Poisson equations}

The system of Vlasov--Poisson equations for describing cosmological dynamics
in a system of $N$ particles of equal masses $m_i=m$
(stars, galaxies, ...) can be represented as follows:
\begin{equation}
\frac{\partial F({\bf x},{\bf v},t)}{\partial t} +
{\rm{div}}_{\bf x}({\bf v}F) +\widehat{G}(F; F) =0,
\end{equation}
$$
\widehat{G}(F; F) \equiv  -{\rm{div}}_{\bf v}\bigg(m^{-1}
\big({\nabla}_{\bf x}(\Phi (F) \big)F \bigg),
$$
\begin{equation}
{\Delta}_{\bf x}^{(d)}\Phi(F)\big|_{t=t_*, \forall{t_*}\in
\mathbb{T}}= 4\pi G  m \int F({\bf x},{\bf v},t_*)\:d{\bf v},
\end{equation}
\noindent
where $F({\bf x},{\bf v},t)=NF_1({\bf x},{\bf v},t)$.
The third term on the right-hand side of this kinetic equation can be represented as a $d$--dimensional ``source--like'' form, where
\begin{equation}
\widehat{G}(F; F) =  m^{-1}{\bf G}(F)\frac{\partial F}{\partial {\bf v}},~~~
\end{equation}
\begin{equation}
{\bf G}(F)=-\nabla_{\bf x}\int\int  {\mathfrak Y}_d^{(\vartheta)}
({\bf x}-{\bf x}')  F({\bf x}',{\bf v}',t_*)\:d{\bf x}'d{\bf v}',
\end{equation}
$$
{\mathfrak Y}_3^{(\vartheta)} ({\bf x}-{\bf x}')=
\frac{G m}{|{\bf x}-{\bf x}'|}+\frac{\vartheta}{6}c^2\Lambda^2 |{\bf x}-{\bf x}'|^2,
$$
\noindent
where $\vartheta \in \{ 0;\,1\}$ and $\vartheta =1$ corresponds to taking into account of repulsive $\Lambda >0$.

The Newtonian potential $\Phi_{N}(r)=-\gamma m/r$ increases
on the interval $r\in (0,+\infty)$ ($\Phi_N \in (-\infty,0)$),
while the potential of the second term of Eq.(1) (see Fig.1),
\begin{equation}
\Phi_{GN}(r)\equiv -{\mathfrak Y}_3^{(1)}(r)
,\end{equation}
has maximum
\begin{equation}
\Phi_{GN}^{(max)}= -\frac{1}{2}{3^{2/3}}(G m c)^{2/3}\Lambda^{1/3}
\end{equation}
at
\begin{equation}
r_c = \big(3G m/\Lambda c^2\big)^{1/3}.
\end{equation}

As shown in \cite{GS7}, for example for the parameters of the Virgo Supercluster, $r_c$ yields around 13 Mpc.

\begin{figure}[h]
\centering
\includegraphics[height=50mm,width=100mm,angle=0]{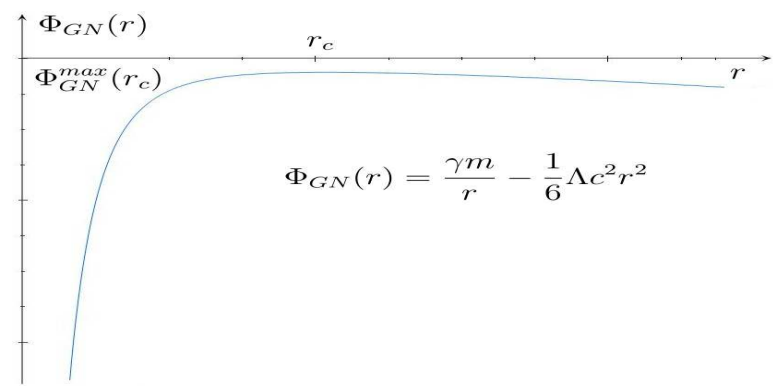}
\vskip 4pt
\caption{\small Modified gravitational potential $\Phi_{GN} ({r})$.}
\label{fig.1}
\end{figure}
\vskip 4pt

The Poisson equation (3) takes the form of the Liouville--Gelfand equation
\cite{Gelfand} if we take the density particle distribution
determined by Maxwell--Boltzmann distribution functions
$F_0({\bf x},{\bf v},t_*)\propto \Omega
\big(\varepsilon[{\bf v}(t_*);\Phi_{min} ({\bf x})]\big|_{ext} \big)$ in terms of velocities.

In this case, in the energy neighborhood of the local extremum
\begin{equation}
d\Omega/d\varepsilon\big|_{\Phi_{min}\to\bar{\Phi}_{min}(r_c)}\approx 0
,\end{equation}
\noindent
we get the relative equilibrium solution for system Eqs.(2)--(3); it is characterized,
in addition to the distinguished energy level of particles on the regular branch of the potential
(in this case, the force term in the Vlasov equation is either canceled or
is the turning point on the phase plane for the evolution of the system),
also the average temperature of the system
$T$ and potential gauge or density $\rho_0=A\exp(-\Phi (0)/T)$
spatial distribution in the  point are the following  
\begin{equation}
F_{0}({\bf x},{\bf v})\big|_{t=t_*}=
B\exp\bigg(  -\frac{m{\bf v}^2}{2T}-\frac{\Phi}{T}+a_j {\mathcal I}_j
 \bigg)\bigg|_{\Phi=\Phi_{min}},
\end{equation}
$$
B \equiv  N \rho_0 \bigg(  \frac{m}{2\pi T}
 \bigg)^{3/2}.
$$

In addition to the energy, the arguments of the function are solutions to the first equation of the $\Omega$ system
of other integrals' motion of ${\mathcal I}_j$, but no changes in this case develop, and one
can consider ${\mathcal I}_j\equiv 0$.

Passing to the density via the integration of the distribution function over
velocities, we obtain the Poisson equation (3) in the form
\begin{equation}
\Delta^{(3)}\Phi = 4\pi G B m
\exp\bigg( -\frac{\Phi}{T}\bigg)\int\int\int
\exp\bigg( -\frac{m{\bf v}^2}{2T} \bigg)d{\bf v} 
\end{equation}
$$
 =4\pi G mN\rho_0 \exp\bigg( -\frac{\Phi}{T}\bigg).
$$
\noindent

Renaming $U=-\Phi/T$, $\lambda = 4\pi \gamma mN
\rho_0 T^{-1}$, we obtain an equation in the form $\Delta^{(3)}U+\lambda \exp(U)=0$.
In what follows, as the gravitational potential, we use 
the function $\Phi({\bf x})$ and the reduced function $U({\bf x})$.
According to Gidas--Ni--Ni\-ren\-berg \cite{Gidas}
all solutions of the last equation are radially symmetric.
Therefore, we need to consider the solutions for
sustainable branches
$U_\lambda (r)$ of the equation
\begin{equation}
U_{rr}''+(2/r)U_r'+\lambda \exp(U)=0
\end{equation}
and corresponding to them in Eq.(6)
radial-symmetric functions
\begin{equation}
\Phi_\lambda (r)=
-TU(r),~~~{\| U\|}(\lambda)~~ {\mbox{is}}~~{\mathcal{J}}-{\mbox{valued\,function}},
\end{equation}
\begin{equation}
{\mathcal{J}}\big|_{\lambda \in (0,2)\cup(2,\:3.32)}\in \{ 1,2,...,
{\mathcal{J}}_{max} \},~~ {\mathcal{J}}\big|_{\lambda=2}=\infty,
\end{equation}
$$
{\mathcal{J}}\big|_{\lambda=3.32}=1
$$
(for $\lambda\gtrsim 3.32$ there are no regular solutions of the Liouville-Gelfand equation).

We are interested in a qualitative analysis and description of the properties of solutions of a linearized
Vlasov--Liouville--Gelfand system, leading
to the possible realization of pseudo--ordered structures.
Thus, we assume that the transfer processes proceed slowly enough, so that an unbalanced particle system allows for the use of a quasi--equilibrium
microcanonical ensemble.

We consider the case of a $d$-dimensional dynamical system ($d=2$ and $3$) and
the linearized version of the Vlasov equation; in this case, this
means that the dynamics
of the system of gravitating particles is considered with respect to the background of its conditional equilibrium
state that changes over time in a known way, in other words
autonomously, on the full iso--energetic surface of the system.

Thus, we arrive at a linear integro--differential equation, as in $(2d+1)$--dimensional
phase space ${\mathbb R}_{\bf x}^d\oplus
{\mathbb R}_{\bf v}^d\oplus {\mathbb T}$) reduced to the integral
equation (over ${\mathbb T}\subseteq {\mathbb R}_t^1$).
We represent its solution in the following form: $F({\bf x},{\bf v},t)=
F_{0}({\bf x},{\bf v}) + f({\bf x},{\bf v},t)$,
$|F_{0}|\gg |f|$. Substituting this expression into the Vlasov equation, we obtain
\begin{equation}
\frac{\partial f}{\partial t}+ {\bf v}\frac{\partial f}{\partial {\bf x}}=
\frac{\partial F_{0}({\bf x},{\bf v})}{\partial {\bf v}}
\cdot
\frac{\partial}{\partial {\bf x}}\int\int {\mathfrak Y}_3^{(\vartheta)}
(|{\bf x}-{\bf x}'|)
f({\bf x}',{\bf v}',t)\:d{\bf v}'d{\bf x}'.
\end{equation}
Using the inversion of the total derivative operator (as semi-groups of
linear operators),
we arrive at the form of the Vlasov equation in the representation of shifts by
trajectories
\begin{equation}
f({\bf x},\:{\bf v},\:t)= f \big({\bf x}-{\bf v}(t-t_0),\:{\bf v}, \:t_0 \big) +
\end{equation}
$$
+ \int^t_{t_0} dt'\: \tilde{B}
\nabla_{\bf v}\exp\bigg(-\frac{m{\bf v}^2}{2T} -\frac{1}{T}\Phi_{-}(t-t') \bigg )
\cdot\widehat{\mathcal{K}}_d^{(-)}[\vartheta]\varrho,
$$
\begin{equation}
\Phi_{-}(t-t') \equiv \Phi_{min}({\bf x}-{\bf v}(t-t')),
\end{equation}
\begin{equation}
\widehat{\mathcal{K}}_d^{(-)}[\vartheta]\varrho \equiv
\frac{\partial}{\partial {\bf x}}\int
{{\mathfrak Y}}_d^{(\vartheta)} (|{\bf x}-{\bf x}'-{\bf v}(t-t')|)
\varrho({\bf x}',t')\:d{\bf x}'.
\end{equation}
Here, $\varrho({\bf x}',t')\equiv
\int f({\bf x}',{\bf v}',t')\:d{\bf v}'$ and $t_0$ certain initial moment,
allocated on the time axis, for example, at a perturbation in
a system described by retarded potentials, $\tilde{B}=B/m$.

To simplify the calculations, we
 assume the temperature to be constant in that part of the system in which
we are interested; if we introduce a temperature field, then it is also necessary
to take the shift of the argument of this field into account when
transitioning to Eq.(16) (the presence of a nonconstant
temperature field can modify the obtained results).

The solution $f({\bf x},{\bf v},t)$ is sought in the form of integrals of
Fourier--Laplace type  from spatial harmonics'
waves with decreasing amplitude (or waves with gaps in the coefficients). Then
it is in the form $\int_{{\mathbb R}^3_{\bf w}} f_{\bf w} ({\bf v},t)
\exp(i {\bf w}{{\bf x}})d{\bf w}$,
${\bf w}=({w}_\ell)^T_{\ell=1,2,3}$,
$w_{\ell}= w_{\ell}^{(Re)}+i w_{\ell}^{(Im)}$
 (for the density $\varrho ({\bf x},t),$ we have
$\varrho ({\bf x},t) =\int_{{\mathbb R}^3_{\bf w}}
 \varrho_{\bf w} (t)\exp(i {\bf w}{\bf x})$)d{\bf w}).

The kernel on the right-hand side of the definition of $\widehat{\mathcal{K}}_3\varrho$, together with the exponentially decreasing part of
 density, can become an integrable
 function (in a limited region of space
 ${\mathbb R}^3$ due to the ``infinite mass problem''), obviously, when the appropriate conditions are met for imaginary 
 indicators' exponents, $w_{\ell}^{(Im)}$. The elements $f_{\bf w}$ and $\varrho_{\bf w}$ belong to a
 set of coefficients of the generalized integral
 Fourier \cite{Tit}, and the existence conditions for these integrals a priori lack an exponential growth
 distribution function $f({\bf x},\:{\bf v},\:t)$ (as we see below, one can
 deduce the quite obvious, albeit somewhat cumbersome,
  criterion for this fact).

Before moving on to Fourier--images, we must integrate Eq.(16) in terms of velocities, to form an equation for density.
On its left-hand side is the Fourier coefficient $\varrho_{\bf w}(t)$, and the first
term on the right--hand side of the resulting integral equation
has the form
\begin{equation}
{}^{[1]}{\mathcal D}f_{\bf w}(t)=\int f_{\bf w}({\bf v},t_0)
\exp\big( - i{\bf w}{\bf v}(t-t_0) \big)\:d{\bf v}.
\end{equation}
\noindent
The second term on the right-hand side of Eq.(16) can be converted to
$\int^t_{t_0} {}^{[2]}{\mathcal D}_{\bf w}(t-t')dt'$, where
\begin{equation}
{}^{[2]}{\mathcal D}_{\bf w}(t-t')=
\int_{{\mathbb R}^3_{\bf v}}d{\bf v}
\bigg[ G \tilde{B} \nabla_{\bf v}
\end{equation}
$$
\exp\bigg(-\frac{m{\bf v}^2}{2T}
-\frac{\Phi_{-}(t-t')}{T} \bigg)\bigg]_1
\exp\big(i{\bf w}({\bf x}- {\bf v}(t-t') )\big)\times
$$
$$
\times
 \frac{\partial}{\partial {\bf x}} \bigg[\int d{\bf x}'\:
{|{\bf x}-   {\bf v}(t-t') - {\bf x}'|^{-1}}
$$
$$
\exp\big(-i{\bf w}({\bf x}-   {\bf v}(t-t') - {\bf x}')\big)
\varrho({\bf x}',t')\bigg]_2.
$$

It should be noted that, in the vicinity of the point $\Phi=\bar\Phi_{min}$ in the phase
plane, the fixed branch of the function of the extrema of the gravitational
potential is close to constant or has a locally parabolic structure
(which corresponds to the motion of a particle
near a state of stability at the generalized Lagrangian point).

The change in the temperature parameter can lead to a change in the
multivalued potential branch and the loss of the characteristics of a singular
point of the function $\Phi_{min}$. As a consequence, among other things,
the impossibility of a significant difference in the velocity dispersion in
structures can, in principle, be explained via the Vlasov--Poisson formalism:
their ordering is controlled by the local
temperature of the system, the change of which leads to a nonsmooth,
jump-like bifurcation of the potential branch, with the inevitable displacement of
libration points and, accordingly, a possible restructuring of the
existing structure.

Renaming
$|{\bf x}- {\bf v}(t-t') - {\bf x}'|=\xi$ and integrating $[...]_2$ in
brackets, we get
\begin{equation}
4\pi \int^\infty_0 {\exp(w^{Im}\xi)}\frac{\sin(w^{Re}\xi)}{w^{Re}\xi}\xi\:d \xi
\equiv {\mathcal H}(w^{Re},w^{Im}),\end{equation}
and finally
\begin{equation}
{}^{[2]}{\mathcal D}_{\bf w}(t-t')=
\end{equation}
$$
{\mathcal H}(w^{Re},w^{Im})
\int_{{\mathbb R}^3_{\bf v}}d{\bf v}
\exp\big(i{\bf w}({\bf x}- {\bf v}(t-t') )\big)
 i\bigg({\bf w}\cdot\bigg[... \bigg]_1\bigg).
$$
Thus, to obtain the Fourier transforms of the density
$\varrho_{\bf w}$, we use the
Volterra (second kind) integral equation
\begin{equation}
\varrho_{\bf w}(t) ={}^{[1]}{\mathcal D}f_{\bf w}(t) +
\int^t_{t_0} {}^{[2]}{\mathcal D}_{\bf w}(t-t') \varrho_{\bf w}(t')\:dt'.
\end{equation}
The solution for this type of equation can be obtained in
terms of the one--sided Laplace transform (by the time variable)
\begin{equation}
\widehat{\mathcal{L}}f_{\bf w}(\omega)=
\int^\infty_0 f_{\bf w}(t)\exp(-\omega t)dt,
\end{equation}
$$
f_{\bf w}(t) =
Heav(t)(2\pi i)^{-1}\int^{\zeta_0 +i\infty}_{{\zeta_0 +i\infty}}
\widehat{\mathcal{L}}f_{\bf w}(\omega) \exp(\omega t)d\omega,
$$
\begin{equation}
\varrho_{\bf w}(t) =(2\pi i)^{-1}
\int^{\zeta_0 +i\infty}_{\zeta_0 -i\infty}\exp(\omega t)
\frac{\widehat{\mathcal{L}}({}^{[1]}{\mathcal D}f_{\bf w}\big)\big(\omega)}{1 -
\widehat{\mathcal{L}}\big({}^{[2]}{\mathcal D}_{\bf w}\big)(\omega)}\:d\omega,
\end{equation}
$$
 \widehat{\mathcal{L}}\big({}^{[k]}{\mathcal D}_{\bf w}\big)(\omega)=
 \int^\infty_0\exp(-\omega t){}^{[k]}{\mathcal D}_{\bf w}(t)\:dt.
$$
\noindent
The integral is taken in the right half--plane of the complex plane
$\omega = -\zeta-i\infty$, along a line parallel to
axes ${\Re}(\omega)=0$ and ${\it Heav} (t)$, where is the Heaviside function.
The poles of the integrand are determined by the equation
$\widehat{\mathcal{L}}\big({}^{[2]}{\mathcal D}_{\bf w}\big)=1$,
which is reduced to the form (for ${\Re}(\omega)>0$)
\begin{equation}
\gamma {\mathcal H}(w^{Re}_\ell,w^{Im}_\ell)
\int_{\mathbb{R}^3}\frac{i {\bf w}\nabla_{\bf v}F_0}{\omega+i {\bf w}{\bf v}}=-1.
\end{equation}
This is the desired dispersion equation in the general case, which, of course,
can only be solved numerically. However, in the limiting cases
$\omega=\omega ({\bf w})$, it is fairly easy to determine
using asymptotic estimates. In particular, for the small
wave-vector  modulus,
we can  expand it in a Taylor series of $|{\bf w}|$ (up to the first
nonvanishing term) the factor in the integrand
($\omega/(\omega +i{\bf v}{\bf w})\approx {\bf w}/\omega -
i{\bf v}{\bf w}^2/\omega^{2}$),
and choose the
coordinate system with the abscissa axis aligned with the vector
${\bf w}$ (to get a scalar dependency).

Then
\begin{equation}
\gamma {\mathcal H}({\bf w})\frac{{\bf w}^2}{\omega^2}
\int \bigg[BT^{-1}\big(-mv_1+\Phi_{min}'\cdot(t-t')\big)
\end{equation}
$$
\exp\bigg(-\frac{m{v_1}^2}{2T} -\frac{\Phi_{-}(t-t')}{T} \bigg)\bigg]_3\:d{v_1 }=1.
$$
Thus, we obtain an approximate dispersion relation
(integrated over time), which can be briefly written as
\begin{equation}
\omega^2 = \gamma {\bf w}^2{\mathcal H}({\bf w})\int\big[...\big]_3dv_1.
\end{equation}
The quantity ${\mathcal H}(w^{Re},w^{Im})>0$, when the dynamical
system depends on the sign of the expression in brackets $[...]_3$,
if $[...]_3<0$, then, $\omega \in {\mathbb C}$ (a purely imaginary number),
otherwise, $\omega$ is real. This means
that the perturbation in our system,
$\sim \exp(\omega t + i {\bf w}{\bf v}),$ will have
an oscillatory
character in time if the system  is in a conditional equilibrium at
high wavelength and when criterion (28) is satisfied
for imaginary $\omega$. For real frequencies, the primary perturbation
is fading or growing (unstable) depending on the sign
$\omega\lessgtr 0$. In addition, the integrand
can change the sign, passing from the oscillatory behavior of the system
to the osculatory one; this is affected by the increase in
observation time, as well as at the transition to another branch of the multivalued function $\Phi_{min}$.

For the case of the $d=2$ system, the situation is
simplified radically.
This is due to the nature of the solutions of the Liouville--Gelfand equation (LGE) in the
two-dimensional case: the number of solution branches
varies depending on the $\lambda$ parameter (the same as
in the three-dimensional case), but in that case it
is either two (for $\lambda < \lambda_{crit}$) or zero --- when
$\lambda > \lambda_{crit}$ and there are no regular solutions.

For the critical value of the solution parameter of LGE,
one can consider the corresponding pair $(\lambda_{crit}, \Phi_{crit})$
as a singular point of the potential; however,
the set of changes in the $\lambda$ parameter do not always include the critical value.
It is essential that we know the explicit form of the LGE solutions to
therefore be able to move away from the local description of the system: neighborhoods of
libration points
in the three-dimensional system are spatially separated
areas, and the above mathematical apparatus
mainly focused on the application of limited
systems. Naturally, one can move away from the mechanical
equilibrium and turn to the thermodynamic description of the state
systems; however, in an explicit form, this is extremely time consuming due to
writing the potential in implicit representation and its
near--equilibrium thermodynamics. The latter can make the
analysis of physical applications of the kinetics of the system difficult at significant
deviations from the initial position taken as equilibrium one.
However, for the two-dimensional system, the situation is different: as previously mentioned,
the gravitational potential of the $N$--particle system can be represented
in a visible form, and we actually always use its uniqueness
(the minimal solution of LGE).


The nonlinear Poisson equation (6) for $d=2$ has the form
\begin{equation}
\Delta^{(2)}\phi({\bf x}) = 4\pi G_2 m^2 N \exp\big( - \phi({\bf x})/T \big),
\end{equation}
$$
\phi({\bf x})=2G_2 m^2 N\int F({\bf x}',{\bf v}')
\ln\big( |{\bf x}-{\bf x}'|/{\mathfrak L} \big)d{\bf x}'d{\bf v}'.
$$
\noindent
In this case, the asymptotic condition is
$\lim_{|{\bf x}|\to\infty}|\phi({\bf x})-2G_2 Nm^2 \ln\:|{\bf x}|)|=0$.
Here $G_2$ is the gravitational constant of
the two--dimensional model, and ${\mathfrak L}$ is the dimensionless
factor, which we omit below. If you apply this
to the model representation of our system
isolated from external mass flows and preserve the
global temperature regime (in the three--dimensional case, we
imply a similar situation), then
in determining
the conditional extremum
($N=const_1$, $\sum\varepsilon_k=const_2$),
the Gibbs entropy $S(F)$ of the system on the plane in
closed region $\bar\Omega$
(${\bf x}\in \bar\Omega$, $\bar\Omega:\:|{\bf x}|\leqslant R$)
corresponding to the maximum of $S(F)$
is the particle distribution function
\begin{equation}
F_{m}({\bf x},{\bf v})=\frac{a_1}{a_2+a_3\cdot|{\bf x}|^2}\exp\big( -a_4 {\bf v }^2 \big)
\xrightarrow{R\to\infty}
\end{equation}
\begin{equation}
\frac{\exp\big( {2E}/{N^2}-2 \big)}{\pi^2}
 \big( \exp(2E/N^2 -2)+{\bf x}^2 \big)^{-1}\exp(-{\bf v}^2/N),
\end{equation}

 $$
E=\frac{N}{2}\int\big({m{\bf v}^2}+
\phi ({\bf x})\big)F({\bf x},{\bf v})d{\bf x}d{\bf v},
$$
\begin{equation}
a_1 = \frac{G_2 m^2 N/T}{2\pi^2}
(2-G_2 m^2 N/T)\frac{R^2}{2},~~~a_2=(2-\gamma_2 m^2 N/T)\frac{R^2}{2} ,\end{equation}
$$
a_3=G_2 m^2 N/2T,~~~
a_4=\frac{m}{2T}.
$$
\noindent
Such a minimal state is realized (uniquely) only if $T>T_c$, $T_c= G_2 m^2N /2$.

Thus, substituting the above formulas into the system of Eqs.(2)--(3), taking the replacement in (4) of the integral
kernels $Y_3\to Y_2({\bf x},{\bf x}')=2G_2 m\ln(|{\bf x}-{\bf x}'|)$ into account,
we can repeat all the calculations of this section for the two--dimensional case,
taking 
the Maxwell--Boltzmann distribution $F_{m}({\bf x},{\bf v}) $ for a conditionally equilibrium state. This allows us
to perform explicit differentiation in square brackets $[...]_1$ in
the definition of ${}^{[2]}{\mathcal D}_{\bf w}(t-t')$.
This leads to the possibility of explicitly observing the change of sign in
brackets $[...]_3$ in the integrand on the left-hand side
dispersion relation
\begin{equation}
-\frac{\Pi_1 \big([a_k],v_1^j,\Delta t\big)}{\Pi_2 \big([a_i],v_1,\Delta t\big)}\bigg|_{i,k=1,...4;\:j=1,...3}\times a_1 \exp\big( a_4 v_1^2 \big) \equiv {\mathfrak D}
,\end{equation}
\noindent
where $\Pi_{1,2}\big( [a_k],v_1^j,\Delta t  \big)$
are polynomials in variable $v_1$ (depending on parameters $a_k,\:\Delta t$).

We emphasize a significant difference from the case $d=3$: we are now considering
the only minimal branch of the inter-particle gravitational potential;
therefore, a slight change in the   parameter norms  does not lead to
a qualitative change in the behavior of the system (far from the singular points'
dispersion equation). At the same time, for three--dimensional
gravitational potential under conditions of polysemy of the function
$\Phi_{min}$ for a fixed $\lambda$, in principle,
a situation of branch inversion arises (the potential jumps
move to a new stable position corresponding to
a more energetically favorable state of the system),
when the norm of the particle distribution function increases
(in nonstationary systems with matter inflow). In addition,
when  we use the Fourier--Laplace transform in the three-dimensional case,
we can see that the
values of $w^{Im}$ are allowed only in the lower half--plane.
Otherwise, for the convergence of the integral on the
infinite interval
in the expression for ${\mathcal H}({\bf w})$ (solutions are only equivalent to
damped density waves in space), and for
two--dimensional systems, it is enough to use the usual Fourier transform. In this case
the size of the system can be arbitrarily large  and the periodicity of the system is violated only
when changing the sign of the integrand
in the dispersion relation.

\section{The unperturbed problem}

In the previous section, we considered the situation when
a system of particles, initially in a conditional equilibrium at some point in time, undergoes a perturbation
and the explicit expression was obtained for the dispersion relation in
the case of long-wave perturbations.
It is possible to analyze the unperturbed case in infinite or semi-infinite
time intervals. At the same time, however, the
conditions for the formation of inhomogeneities in space
and the solution of the Vlasov--Poisson system becomes
``quasi-discrete.'' Namely,  randomly distributed,
intersecting one-dimensional ``channels'' do arise, consisting
bunches of particles separated by voids, as
the intersecting three-dimensional structure of the cosmic web (\cite{web}).

Let us turn to a particular solution of the Vlasov--Poisson equation
(for the three-dimensional case) of the form
\begin{equation}
f^{+}({\bf x},{\bf v},t) \sim c(k) f^\dag_{\bf w} ({\bf v})
\exp(i\omega t -i {\bf w}{\bf x}).
\end{equation}
\noindent

Let us take as a simplifying assumption that, in the expression
\begin{equation}
F_0({\bf x},{\bf v})=B \exp\bigg(-\frac{m{\bf v}^2}{2T}
-\frac{1}{T}\Phi_{min}({\bf x})\bigg)
,\end{equation}
the norm of quantity $\Phi_{min}\approx \varphi = {\rm const}$, that is when the state of the system is close to relative equilibrium, in which
by definition $\Phi_{min}'=0$.

We, as before, follow the analysis methodology proposed in \cite{V1}. As a result,
we have\begin{equation}
f^\dag_{\bf w} ({\bf v}) =
\frac{G}{m} {\mathcal H}({\bf w})\frac{({\bf w}
\nabla_{\bf v}{\mathfrak{M}}({\bf v}) )}{{\bf w}{\bf v}-\omega}
\int f^\dag_{\bf w} ({\bf v})dv_1dv_2dv_3,
\end{equation}
$$
{\mathfrak{M}}({\bf v})
= B\exp(-\varphi/T)\exp(-m{\bf v}^2/2).
$$
The condition for the existence of a nontrivial solution as of functions
$f^{+}$ is obtained after integrating over the velocities of both parts
(dispersion relation $\omega = \omega ({\bf w})$)
\begin{equation}
G m^{-1} {\mathcal H}({\bf w})B\exp(-\varphi/T)
\int \frac{({\bf w}{\rm grad}_{\bf v})}{{\bf w}{\bf v}-\omega}d{\bf v}=1.
\end{equation}
\noindent
The values of the components of the wave vector ${\bf w}$ are
complex and the integral is taken in the sense of the principal value symmetrically along the axis $v_1$,
 if the $w$ axis is along the axis of the radial coordinate. The sum of two terms when going around the pole corresponds to the retarded and leading potentials.

Hence, we have
\begin{equation}
f^\dag_{\bf w}({\bf v}) = c({\bf w})
G m^{-1} {\mathcal H}({\bf w})B\exp(-\varphi/T)
({\bf w}\nabla_{\bf v}/(\omega-{\bf w}{\bf v})).
\end{equation}

One can obtain an analytical representation of the dispersion relation
if we approximate the Maxwellian ${\mathfrak M}$
\begin{equation}
\lim_{|{\bf v}|\ll \infty}{\mathfrak M} \to N\sqrt{2T/m}/\pi
\big( 2T/m+v_1^2\big)^{-1}\equiv{\mathfrak M}^\dag.
\end{equation}
Then
\begin{equation}
\int dv_1 \frac{d{\mathfrak M}^\dag/dv_1}{m(v_1-\omega/w)}=
-\frac{N}{\pi T}
 \frac{\pi}{2}\frac{1-\nu^2}{(1+\nu^2)^2},~~~\nu =\omega/w \sqrt{m/(2T) }.
\end{equation}
Therefore, the dispersion relation takes the form
\begin{equation}
\bigg( \frac{4\pi}{m}B\int^\infty_0
{\exp(w^{Im}\xi)}\frac{\sin(w^{Re}\xi)}{w^{Re}\xi}\xi\:d\xi \bigg)^{- 1}=
-\frac{1-\nu^2}{(1+\nu^2)^2}.
\end{equation}

Similarly, one can obtain the dispersion relation
for any function ${\mathfrak M}(v_1)$.
So, we got a $\omega ({\bf w})$ dependency for the system of particles with self-gravity.
This provides information about the nature of the dynamics of the system, that is it is
oscillatory in time and/or space with decreasing or conserving amplitude being exponentially decreasing and unstable.
 
 The importance of the analysis of dispersion solutions lie in the fact that no random fluctuations have a role in the
mechanism of emergence of pseudo-oscillatory (in time and in selected spatial directions) structures of matter. While explicitly expressed spatial periodicity of structures does not occur, the dispersion relations reveal certain ordering (sequencing) of the structures. 
 
 Thus, both the emergence of macro-structures, and their distribution
 is immanently inherent in the technique using the Vlasov-Poisson equations at the potential of Eq.(1).

\section{Conclusions}

 The emergence of two-dimensional structures as Zeldovich pancakes is currently
associated with density perturbations, whose growth is described by
 classical or weakly relativistic hydrodynamics, without considering the role of gravity.
 Then,  certainly the problem is to ensure the emergence of the needed perturbations.

The issue we dealt with in this study is whether a kinetic description -- taking the role of self-consistent gravity of the
 many-particle ensemble into account, that is to say within the formalism of the Vlasov equation -- can also lead to similar filamentary structures.

Within the Vlasov formalism \cite{V1,V2}, we analyzed a kinetic model and revealed the occurrence of semi-periodicity of the arisen 
structures -- voids -- separated by two-dimensional surfaces (walls). On the one hand, the emergence of  two-dimensional structures have clear differences from the hydrodynamical approach, since is not depending on the form of the initial perturbations, but it is due to 
the self-consistent gravitational interaction only. On the other hand, the emergence of  two-dimensional  structures (walls) in the three-dimensional N-body problem is not trivial a priori.

The crucial point in our kinetic approach is the use of the potential of Eq.(1) with the repulsive cosmological $\Lambda$ term. 
The latter does not include any free parameters (e.g., scalar fields, etc); however, as outlined above, it is based on one 
of the very first principles, the generalized form of a potential  derived in \cite{G1} to ensure an identical 
gravity for the sphere and point mass. As previously stated, this potential already has certain empirical bases, enabling one to describe 
the dynamics of galaxy clusters, and the Hubble tension as a result of local and global flows of galaxies. We now 
see that the same approach within the Vlasov formalism is able to predict certain features of the cosmic structures, for example of voids. 

Of course, the important point for any theory or model is the study of the correspondence -- qualitative and whenever possible, quantitative -- of the predictions with the observations, as for example the analysis in \citet{Bo} of the predictions of the nonlocal gravity model with gravitational lensing observations of the CLASH survey (\cite{Po}).

In our case,{ qualitative} correspondence can be considered as the prediction of two-dimensional structures (walls) and voids between them, obtained at the rigorous analysis of the three-dimensional system. This nontrivial result is due to the self-consistent interaction being considered which has both attractive and repulsive terms.    

As for the {quantitative} criterion, we have derived the scale (size) of the voids, namely, that scale is determined by the 
critical radius $r_c = \big(3G m/\Lambda c^2\big)^{1/3}$, which is balancing the Newtonian and repulsive terms in Eq.(1).
This enables a direct comparison with observations, for example the size of voids yields around 25 Mpc when using the data of the Virgo Supercluster, which is in agreement with observational data (e.g., \cite{Ce,GK,S2,Na}). As mentioned in \cite{GS7}, the observational data defining
$r_c$ can vary from survey to survey, thus indicating that the numerical scale of the voids can vary depending on the mean matter density of the local region of the Universe. In other words, although the mechanism of the void formation is common -- the interplay between the self-consistent contribution of gravitational attraction and the repulsion with the cosmological constant -- the voids can have a different mean size determined by the local conditions (density) of the Universe. This can become a vast topic for an analysis involving observational surveys.

Our analysis thus indicates the important role that the self-interaction can have in the 
emergence of the voids in the large-scale matter distribution, thus supporting further studies in this direction.

\section{Acknowledgments}

We thank the referee for valuable comments.

\end{document}